\documentclass[11pt,a4paper]{article}
\usepackage{jcappub}

\usepackage{bm}
\usepackage{amsfonts}
\usepackage{subfigure}

\newcommand{\alt}{\mbox{\;\raisebox{.3ex}
  {$<$}$\!\!\!\!\!$\raisebox{-.9ex}{$\sim$}\;}}

\newcommand{\mystrut}{\hbox{\vrule height12pt depth5pt width0pt}}

\renewcommand\[{\left[}

\newcommand{\be}{\begin{equation}}
\newcommand{\ee}{\end{equation}}
\newcommand{\bea}{\begin{eqnarray}}
\newcommand{\eea}{\end{eqnarray}}

\begin{document}

%%%%%%%%%%%%%%%%%%%%%%%%%%%%%%%%%%%%%%%%%%%%%%%%%%%%%%%%%%%%%%%%%%%%%%
% Frontpage %%%%%%%%%%%%%%%%%%%%%%%%%%%%%%%%%%%%%%%%%%%%%%%%%%%%%%%%%%
%%%%%%%%%%%%%%%%%%%%%%%%%%%%%%%%%%%%%%%%%%%%%%%%%%%%%%%%%%%%%%%%%%%%%%

%\subheader{\hfill Preprint ....}

\title{Future cosmological sensitivity for hot~dark matter axions}

\author[a]{Maria Archidiacono,}
\author[a,b]{Tobias Basse,}
\author[c,d]{Jan~Hamann,}
\author[a]{Steen~Hannestad,}
\author[e]{Georg~Raffelt}
\author[b]{and Yvonne Y.~Y.~Wong}

\affiliation[a]{Department of Physics and Astronomy, University of Aarhus,
 DK-8000 Aarhus C, Denmark}

\affiliation[b]{School of Physics, The University of New South Wales,
Sydney NSW 2052, Australia}

\affiliation[c]{Theory Division, Physics Department, CERN,
 CH-1211 Geneva 23, Switzerland}
 
 \affiliation[d]{Sydney Institute for Astronomy, The University of Sydney, Sydney NSW 2006, Australia}

\affiliation[e]{Max-Planck-Institut f{\"u}r Physik (Werner-Heisenberg-Institut),
F{\"o}hringer Ring 6, D-80805 M{\"unchen}, Germany}

\emailAdd{archi@phys.au.dk, tb06@phys.au.dk, jan.hamann@sydney.edu.au, sth@phys.au.dk}
\emailAdd{raffelt@mpp.mpg.de, yvonne.y.wong@unsw.edu.au}

\abstract{We study the potential of a future, large-volume photometric survey to constrain the axion mass~$m_a$  in
the hot dark matter limit. Future surveys such as {\sc Euclid} will have significantly more constraining power
than current observations for hot dark matter. Nonetheless, the lowest accessible axion masses are limited by the fact that axions lighter than
$\sim 0.15$~eV decouple before the QCD epoch, assumed here to occur at a temperature $T_{\rm QCD} \sim 170$~MeV; this leaves an axion population of such low density that its
late-time cosmological impact is negligible.  For larger axion masses, $m_a \gtrsim 0.15$~eV, where axions remain in equilibrium
until after the QCD phase transition, we find that a {\sc Euclid}-like survey combined with Planck  CMB data  can detect $m_a$ at very high significance. 
Our conclusions are robust against assumptions about prior knowledge of the neutrino mass. 
Given that the proposed IAXO solar axion search is sensitive to
$m_a\lesssim0.2$~eV, the axion mass range probed by cosmology is nicely
complementary.}

\maketitle

\section{Introduction}

Interest in the low-energy frontier of particle physics has surged over the past few years, especially in the
area of axions and axion-like particles (ALPs)~\cite{Jaeckel:2010ni, Hewett:2012ns, Essig:2013lka,
Baker:2013zta}. The ADMX microwave cavity experiment is getting ready for its final push towards covering a
well-motivated range of axion masses at the sensitivity required to find these elusive particles if they were
the dark matter in our galaxy~\cite{Asztalos:2009yp, Asztalos:2011bm, vanBibber:2013ssa, Rybka:2014xca,
Cho:2013dwa}. New ideas to broaden the range of search masses are vigorously discussed~\cite{Baker:2011na,
Horns:2012jf, Jaeckel:2013sqa, Jaeckel:2013eha, Horns:2013ira, Sikivie:2013laa, Rybka:2014cya, Graham:2011qk,
Graham:2013gfa, Budker:2013hfa, Beck:2013jha}. In Korea, an entire Center of the Institute for Basic Science,
the Center for Axion and Precision Physics (CAPP), has been dedicated to the search for axion dark matter and
related topics in precision physics~\cite{CAPP}. At DESY, first steps have been taken towards the largest ever
photon-regeneration experiment to search for ALPs~\cite{Bahre:2013ywa}. At CERN, the CAST experiment to search
for solar axions has almost completed its original mission~\cite{Zioutas:2004hi, Andriamonje:2007ew,
Arik:2008mq, Arik:2011rx, Arik:2013nya, Barth:2013sma}, and has paved the way for proposing a much larger
next-generation axion helioscope, the International Axion Observatory~(IAXO)~\cite{Irastorza:2011gs,
Armengaud:2014gea}. In no small amount, our present study is motivated by the IAXO proposal.

Axion helioscopes such as CAST and IAXO take advantage of the two-photo axion interaction vertex both 
as a source of axions in the sun via the Primakoff effect and for the back-conversion of axions into X-rays
in a dipole magnet oriented towards the sun.
However, this back-conversion is not effective if the photon-axion mass difference is too large. 
Therefore, in the $(m_a, g_{a\gamma})$-parameter
plane, where $g_{a \gamma}$ is the two-photon axion coupling strength, 
it has  been traditionally difficult to probe the ``axion line'' on which the
 QCD axion models---as opposed to generic axion-like particles (ALPs)---reside.
CAST has recently succeeded in touching the axion line in a narrow $m_a$ range around 1~eV~\cite{Arik:2013nya}; it 
is the purpose of the proposed IAXO to push to greater sensitivity and
in this way explore realistic axion parameter space. In this connection, we note that, independently of model
dependence of the relationship between $g_{a\gamma}$ and $m_a$, cosmological hot dark matter bounds can
provide a guide as to the largest axion masses that are worth exploring with such an experiment.

Current observations of the cosmic microwave background (CMB) anisotropies and the large-scale matter
distribution already impose severe constraints on the hot dark matter abundance; in the axion interpretation,
this constraint corresponds to an axion mass limit of around 1~eV \cite{Hannestad:2005df, Hannestad:2007dd,
Hannestad:2008js, Hannestad:2010yi, Archidiacono:2013cha, Melchiorri:2007cd}. In the future, large-volume
surveys such as the {\sc Euclid} mission~\cite{Laureijs:2011gra} will have a significantly enhanced
sensitivity to hot dark matter~\cite{Hamann:2012fe,Basse:2013zua}.  The aim of the present work, therefore, is
to examine just what such a survey will be able  to  do for axion physics.

To achieve this goal, we begin in section~\ref{sec:axioncosmology} with a brief review of axion cosmology. In
section~\ref{sec:cosmologicalanalysis} we set up the cosmological framework and in particular discuss the used
data sets and the cosmological model space. The actual numerical analysis for different assumptions about the
hot dark matter contribution of neutrinos is provided in section~\ref{sec:numericalresults}. We 
summarise and discuss our findings in section~\ref{sec:discussion}.

\section{Axion cosmology}
\label{sec:axioncosmology}

\subsection{Hot and cold axion populations}

The primary parameter characterising invisible axion models is the axion decay constant~$f_a$, or,
equivalently, the energy scale at which the Peccei-Quinn symmetry is spontaneously broken. Mixing with the
$\pi^0$-$\eta$-$\eta'$ mesons at low energies induces a mass for the axion, given approximately by
\begin{equation}
m_a=\frac{z^{1/2}}{1+z}\,\frac{f_\pi}{f_a}\,m_\pi=
\frac{0.58~{\rm eV}}{f_a/10^7~{\rm GeV}}\,,
\end{equation}
where $z=m_u/m_d$ is the up/down quark mass ratio.  The value of $z$ lies in the range 0.38--0.58~\cite{Agashe:2014kda}, which leads to an uncertainty in the relationship between $m_a$ and $f_a$ of less than~$\pm5$\%.  Therefore, we may equally use $m_a$ as a characteristic parameter.

\begin{figure}
\centering
\includegraphics[width=0.78\textwidth]{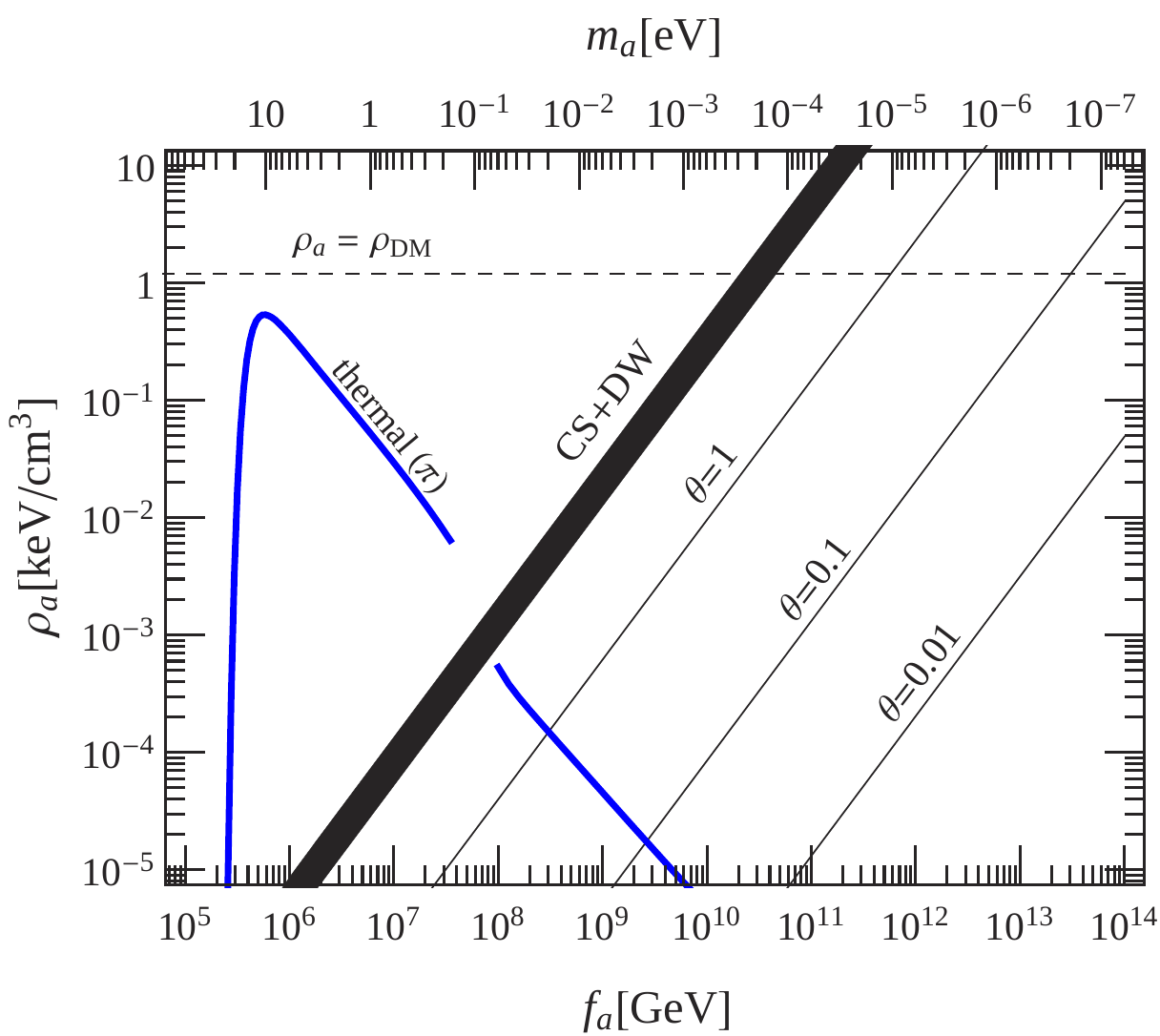}
\caption{Present-day axion dark matter density as a function of $m_a$.  The thermal axion population, which
forms hot dark matter, is represented by the thick blue line. The thick black line denotes the cold axion
population in the scenario in which Peccei-Quinn symmetry breaking occurs after inflation so that the visible
universe contains many patches of different initial axion-field misalignment angles~$\Theta_{\rm i}$; the
energy density shown here subsumes both contributions from the re-alignment mechanism and from cosmic string
(CS) and domain-wall (DW) decay according to reference~\cite{Hiramatsu:2012gg}. The thin black lines pertain
to the case in which Peccei-Quinn symmetry breaking occurs before inflation so that one single initial
misalignment angle~$\Theta_{\rm i}$ pervades the entire visible universe; the cold axion populations for
several different values of $\Theta_{\rm i}$ as indicated at the lines are shown. (Figure adapted from one
supplied by Javier Redondo.)}
\label{fig:axionpopulations}
\end{figure}

Figure~\ref{fig:axionpopulations} shows the ``Lee-Weinberg curve for axions'', pieced together from several
different production mechanisms~\cite{Sikivie:2006ni}. Non-thermal production by the re-alignment mechanism
and possibly by the decay of cosmic strings (CS) and domain walls (DW) produces a nonrelativistic population
that can account for the cosmic cold dark matter.  The thick black line in figure~\ref{fig:axionpopulations}
shows the present-day cold axion energy density and its nominal uncertainty, assuming the Peccei-Quinn
symmetry is broken after inflation so that each causal patch at the QCD epoch has a unique initial
misalignment angle $\Theta_{\rm i}$ when the axion field begins to oscillate. We follow
reference~\cite{Hiramatsu:2012gg} concerning the contribution from strings and domain-wall decay; other
authors find a somewhat smaller contribution~\cite{Sikivie:2006ni, Visinelli:2009zm}.

If the Peccei-Quinn symmetry is broken before inflation such that $\Theta_{\rm i}$ is the same in the entire
observable universe, the axion abundance depends on $\Theta_{\rm i}^2$ and in this sense is a random
number~\cite{Sikivie:2006ni}. Figure~\ref{fig:axionpopulations} shows some examples for the cosmic axion
abundance in this scenario as thin black lines for the indicated $\Theta_{\rm i}$ values. Note also that
because the axion field in this case is present and massless during inflation, the quantum fluctuations it
picks up independently of the inflaton field can later show up as isocurvature fluctuations in the CMB anisotropies unless the energy scale of inflation is sufficiently
small~\cite{Lyth:1989pb, Beltran:2006sq, Hertzberg:2008wr, Hamann:2009yf}. This scenario would have been excluded if
the tensor-to-scalar ratio~$r$ had been as high as initially claimed by the BICEP2 team~\cite{Ade:2014xna}, because 
$r = 0.2$ corresponds to an unacceptably high inflation energy scale
of $\sim 10^{14}$~GeV~\cite{Higaki:2014ooa, Marsh:2014qoa, Visinelli:2014twa}.  However, with the results of the recent combined analysis of
BICEP2, Keck Array and Planck data~\cite{Ade:2015tva} being consistent with $r=0$, reports of its demise may have been a bit premature.

In addition, thermal processes produce axions which today play the role of hot dark
matter~\cite{Turner:1986tb}. The two populations co-exist because thermalisation of the cold population is
extremely slow thanks to the axion derivative interactions with ordinary matter and photons, whose rates are
suppressed at low energies. The thick blue line in figure~\ref{fig:axionpopulations} shows the hot dark matter
density arising from thermal axion production. For axion masses $m_a\gtrsim0.15$~eV, axion-pion interaction is
the main production mechanism~\cite{Chang:1993gm}, and freeze-out occurs after the QCD epoch. However, as
$m_a$ decreases and the freeze-out temperature approaches $T_{\rm QCD}\sim 170$~MeV, the freeze-out epoch
suddenly jumps to a much higher temperature because axion interactions with gluons and quarks before
confinement are much less efficient~\cite{Masso:2002np, Graf:2010tv, Salvio:2013iaa}.  Furthermore, any
thermal axion population frozen out before the QCD epoch is necessarily diluted by entropy production during
the QCD epoch, which causes a sharp drop  in the axion energy density in the transition region at
$m_a\sim0.15$~eV, as shown in figure~\ref{fig:axionpopulations}. The exact transition between the pre- and
post-QCD freeze-out regimes is not known, so we leave a gap in the blue curve. For very large axion masses,
$m_a\gtrsim20$~eV, the axion decay $a\to2\gamma$ is fast relative to the age of the universe, so that the hot
axion population disappears.

The contribution of thermally produced axions to the energy budget of the universe and hence the axion
mass~$m_a$ can be constrained by observations of CMB anisotropies and the large-scale structure distribution
in the same way as we constrain neutrino hot dark matter and hence the neutrino mass sum. Our group has previously
published limits on $m_a$ in a series of papers based on a sequence of cosmological data
releases~\cite{Hannestad:2005df, Hannestad:2007dd, Hannestad:2008js, Hannestad:2010yi, Archidiacono:2013cha};
our most recent limit is $m_a\lesssim 0.67$~eV at 95\% C.L.\ for a minimal cosmological model and using
Planck-era cosmological data~\cite{Archidiacono:2013cha}, a number comparable to the results of other
authors~\cite{Melchiorri:2007cd}. Hot dark matter constraints do not apply to axion masses exceeding
$\sim20$~eV because the aforementioned $a\to 2 \gamma$ decay removes the axion population.  However, we note
that axions masses up to $\sim300$~keV are nonetheless cosmologically forbidden owing primarily to
modifications to the primordial deuterium abundance  instigated by the decay photons~\cite{Cadamuro:2010cz}.

\subsection{Axion decoupling temperature}
\label{sec:dectemp}

The freeze-out history of the axion-pion interaction at temperatures below the QCD phase transition was
studied in detail in reference~\cite{Hannestad:2005df}. This treatment is technically valid for
$m_a\gtrsim0.15$~eV and therefore adequate relative to the sensitivity of the present generation of
cosmological probes because the current upper bound is $m_a \alt 0.67$ eV \cite{Archidiacono:2013cha}.
However, an extension to lower axion masses would be useful in anticipation of the next generation of
cosmological probes. This is especially so in view of the forecasted sensitivity of the {\sc Euclid} mission
to neutrino masses, $\sigma(m_\nu) \sim 0.01$~eV~\cite{Hamann:2012fe}; considering the similar phenomenology,
a decoupling model for axions down to similar masses should be contemplated.

Axions with very small masses decouple at temperatures above the QCD phase transition, where the axion
interaction with free quarks and gluons can be studied perturbatively~\cite{Graf:2010tv}. Perturbative
treatments fail, however, when decoupling occurs close to the QCD epoch, and no full non-perturbative
calculation exists to date. Therefore, to model axion decoupling in the mass range $0.01~{\rm eV}  < m_a < 1~{\rm
eV}$, we use the following prescription.
\begin{enumerate}

\item We fix the QCD phase transition temperature $T_{\rm QCD}$ at a specific value.  A typical choice might be $T_{\rm QCD} = 170$~MeV, which conforms with the
    most recent calculations from lattice gauge theory~\cite{Bazavov:2009zn, Endrodi:2011gv}. The chosen
    value of $T_{\rm QCD}$ does have an impact on the axion mass range that can be probed using large-scale
    structure observation, and we shall return to this point later.

\item For large axion masses ($m_a \gtrsim 0.2$~eV) we continue to use the same decoupling calculation as in
    previous works. We extend this calculation down to an axion mass at which it returns a decoupling
    temperature matching $T_{\rm QCD}$.  For the choice of $T_{\rm QCD} = 170$~MeV, this corresponds to $m_a
    = 0.145$~eV.  For all axion masses smaller than this limit, freeze-out occurs at $T > T_{\rm QCD}$.

\item For all axion masses smaller than this limiting value, we assume an effective number of entropy
    degrees of freedom of $g_*=80$ at the time of freeze-out, independently of the exact axion mass.   This
    approximation is reasonable because $g_*(T)$ varies only logarithmically with temperature from $g_*\sim
    60$ at $T_{\rm QCD}\sim170$~MeV to $g_*\sim 100$ at the electroweak phase transition $T_{\rm EW} \sim
    100$~GeV, in contrast with the step-like variation in the spate of a few MeVs during the QCD phase
    transition (see figure~3 of reference~\cite{Brust:2013ova}).

\end{enumerate}
As we shall see later, axions decoupling at temperatures above $T_{\rm QCD}$ are practically invisible to the
surveys discussed here, whereas axions decoupling after the QCD epoch have a clear impact on cosmological
observations.

\begin{figure}[t]
\centering
\includegraphics[width=0.7\textwidth]{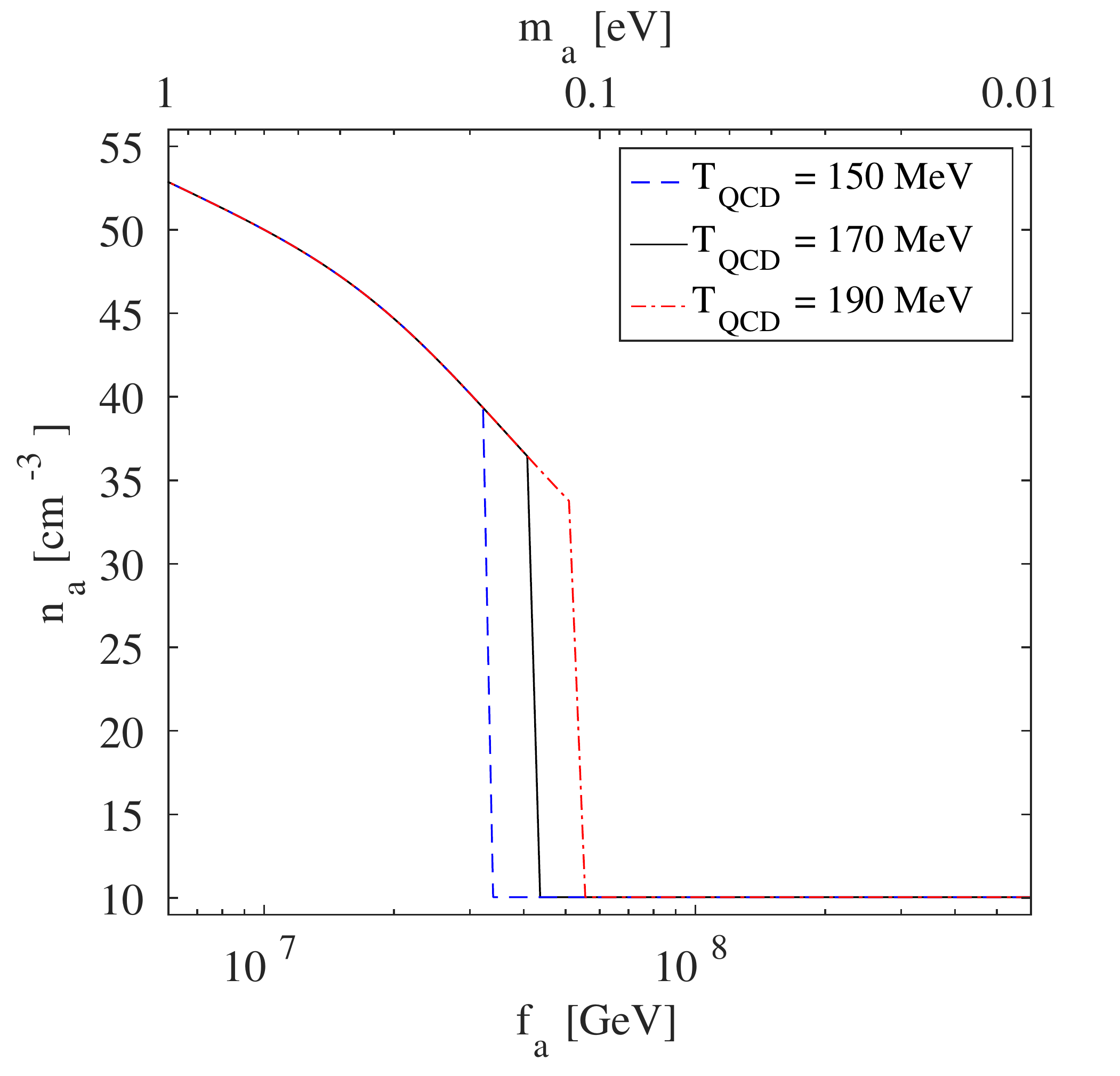}
\caption{Present-day axion number density as a function of the axion mass (upper horizontal axis) and of the
axion decay constant (lower horizontal axis), assuming $T_{\rm QCD} = 170$~MeV (solid black line), $T_{\rm
QCD} = 150$~MeV (dashed blue line) and $T_{\rm QCD} = 190$~MeV (dash-dotted red line).}
\label{fig:na}
\end{figure}

Figure~\ref{fig:na} shows the present-day axion number density as a function of the axion mass according to
the model described above.   For the choice of $T_{\rm QCD} = 170$~MeV, the number density plummets from $n_{a}=36.45\,{\rm
cm}^{-3}$ to $10.04\,{\rm cm}^{-3}$ at $m_a = 0.145$~eV,  reflecting the jump of the axion decoupling temperature when the universe transits through the QCD
epoch. Since the dividing line hinges strongly on the assumed value of
$T_{\rm QCD}$, we show also the cases of  $T_{\rm QCD} = 150$ and 190~MeV, where the corresponding drops in $n_a$
occur at $m_a = 0.185$ and 0.117~eV, respectively. Given the uncertainty in $T_{\rm QCD}$ and the lack of a
full quantitative calculation of the axion freeze-out in this temperature region, we cannot state the
sensitivity to axion masses any more precisely.
Henceforth, we shall use $T_{\rm QCD}=170$~MeV as our benchmark case, and 
\begin{equation}
m_{\rm QCD}=0.145~{\rm eV}
\end{equation}
shall denote the limiting axion mass corresponding to nominal freeze-out at this temperature.

\section{Cosmological set-up}
\label{sec:cosmologicalanalysis}

\subsection{Cosmological mock data}

In order to assess the sensitivity of future cosmological observations to the axion mass we use the
parameter forecast code\footnote{\url{http://jhamann.web.cern.ch/jhamann/simdata/simdata.tar.gz}} first described in reference~\cite{Hamann:2012fe} for galaxy clustering and cosmic
shear surveys, and later extended in reference~\cite{Basse:2013zua} to include cluster measurements.   We
adopt the same assumptions about the observational data, which conform to the specifications of the upcoming
{\sc Euclid} mission \cite{Laureijs:2011gra}.  A summary of the mock data sets is as follows.

\begin{enumerate}

\item The cosmic shear auto-spectrum $C_{\ell,ij}^{\rm ss}$ runs in the range $2 \leq \ell \leq \ell_{\rm
    max}^{\rm s}=2000$, and the indices $i,j \in [1,N_{\rm s}]$ label the redshift bin. As discussed in
    reference~\cite{Hamann:2012fe}, shot-noise always dominates well before $\ell$ reaches $2000$, so that
    the precise value chosen for $\ell_{\rm max}$ is not critical. Redshift slicing is performed in such a
    way that all bins contain similar numbers of source galaxies and hence suffer the same amount of shot
    noise.  We choose $N_{\rm s}= 2$, as no substantial improvement in the parameter sensitivities can be
    expected beyond this point~\cite{Hamann:2012fe}.

\item The galaxy auto-spectrum $C_{\ell,ij}^{\rm gg}$ uses multipole moments running from $\ell=2$ to
    $\ell^{{\rm g},i}_{\rm max}$ in redshift bins $i,j \in [1,N_{\rm g}]$, where $\ell^{{\rm g},i}_{\rm
    max}$ is set following the prescription of reference~\cite{Hamann:2012fe} so as to minimise the
    non-linear contamination.   As shown in the said paper, most cosmological parameter sensitivities saturate
    at around $N_{\rm g}=11$; we shall also adopt this value.  The linear galaxy bias is assumed to be
    perfectly known.  While this assumption is not very realistic, in the absence of a concrete bias model
    the only alternative would be to adopt the opposite extreme in which the bias is completely unknown and
    uncorrelated between redshift bins.  We have tested this equally unrealistic assumption in a previous
    work~\cite{Hamann:2012fe}, and found that the resulting parameter constraints are utterly uncompetitive
    if the galaxy data were used on their own, and, if analysed in combination with shear data, hardly
    better than the shear-only constraints. We therefore do not consider the case of an unknown galaxy bias
    in the present work.

\item The shear--galaxy cross-spectrum $C_{\ell,ij}^{\rm sg}$ in the shear redshift bin $i \in [1,N_{\rm
    s}]$ and galaxy redshift bin $j \in [1,N_{\rm g}]$ runs from $\ell=2$ to $\ell^{{\rm g},j}_{\rm max}$,
    where  $\ell^{{\rm g},j}_{\rm max}$ is determined by the galaxy redshift binning.

\item The cluster mass function measurements~$N_{i,j(i)}$ source from clusters in the redshift range $z \in
    [0.01, z_{\rm high}]$ and with  masses \mbox{$M \in [ M_{\rm thr}(z), 10^{16}~M_\odot]$}, where $M_{\rm
    thr}(z)$ is the redshift-dependent mass detection threshold, and $z_{\rm high}$ the redshift at which
    $M_{\rm thr}(z)$ exceeds $10^{16}~M_\odot$.   The index~$i \in [1,N_z]$ labels the redshift bins, with
    boundaries chosen such that for the fiducial model all bins contain equal numbers of clusters.  In each
    redshift bin~$i$ we further subdivide the cluster sample into mass bins~$j(i) \in [1,N_\mathrm{m}]$,
    again demanding that the same number of clusters should fall into each bin~\cite{Basse:2013zua}.

\item A mock data set from a {\sc Planck}-like CMB measurement is generated according to the procedure of
    reference~\cite{Perotto:2006rj}, together with modifications according to reference~\cite{Basse:2014qqa}
    so that it better captures some of the gross features of the realistic Planck likelihood, such as the
    fact that its effective sky coverage is scale-dependent.

\end{enumerate}

\subsection{Cosmological model space}

We use an 8-parameter cosmological model ${\bm\Theta}$, consisting of six ``vanilla'' parameters ${\bm
\Theta}^{(6)}$, extended by the sum of neutrino masses here denoted as $m_\nu$ and the axion mass $m_a$,
i.e.,
\begin{equation}\label{eq:model}
 {\bm \Theta} \equiv \Bigl({\bm \Theta}^{(6)},m_\nu, m_a\Bigr)
 \equiv \Bigl((\omega_{\rm cdm},\omega_{\rm b},h,
\ln(10^{10}A_{\rm s}),n_{\rm s}),m_\nu,m_a\Bigr).
\end{equation}
Here, $\omega_{\rm cdm} \equiv \Omega_{\rm cdm} h^2$ and $\omega_{\rm b} \equiv \Omega_{\rm b} h^2$ are the
present-day physical CDM and baryon densities respectively, $h \equiv H_0/(100{\rm~km/s/Mpc})$ is the
dimensionless Hubble parameter, $\ln(10^{10}A_{\rm s})$ and $n_{\rm s}$ denote respectively the amplitude and
spectral index of the initial scalar fluctuations, and $z_{\rm re}$ is the reionisation redshift.

For the part of the parameter space not related to neutrinos or axions, our fiducial model is defined by the
parameter values
\begin{equation}
{\bm \Theta}^{(6)}_{\rm fid} = \left(0.0226, 0.1126587, 0.7, 2.1 \times 10^{-9}, 0.96, 11 \right),
\end{equation}
while for $m_\nu$ and $m_a$ we shall test a variety of fiducial values. Specifically, for the axion mass we
use a range of values
\begin{equation}
m_{a{\rm,fid}} = 0.01, 0.1, 0.15, 0.2, 0.5~{\rm eV},
\end{equation}
covering the possibility of axion freeze-out both before and after the QCD phase transition (see
figure~\ref{fig:na}). Neutrinos provide another source of hot dark matter of an unknown size. We expect their
phenomenology to be similar to that of axions so that $m_\nu$ and $m_a$ are likely to be correlated.
Therefore, we vary also the fiducial neutrino mass sum in the range
\begin{equation}
m_{\nu{\rm,fid}} = 0.06, 0.11, 0.3, 0.7, 1~{\rm eV},
\end{equation}
i.e., from the minimal mass allowed in the normal hierarchy by neutrino oscillation experiments, to well
beyond the current cosmological upper bound.

Our analysis assumes, for the most part, a pessimistic scenario in which the neutrino mass sum is an unknown
that has to be fitted simultaneously with the axion mass to the mock data, possibly leading to a degradation
of the sensitivity to the latter. However, we shall consider also an optimistic case in which the neutrino
mass is taken to be infinitely well measured by other means (e.g., by tritium $\beta$-decay experiments);
in practice this means we hold the neutrino mass sum fixed at its fiducial value when fitting the mock data.

Lastly, we note that a number of recent cosmological analyses have found hints for a small hot dark matter
contribution due to either a nonzero neutrino mass sum or axion mass~\cite{Palanque-Delabrouille:2014jca,
Archidiacono:2013cha}. These results are largely a consequence of tension between different observations
within the minimal $\Lambda$CDM interpretation---notably, the conflicting values of $\sigma_8$ and $\Omega_m$
reported by cluster catalogues and by Planck's CMB temperature anisotropy measurements---which tends to settle
on a nonzero hot dark matter component as a middle ground.  However, the tension between these observations
may just as well indicate an incomplete understanding of the systematics~\cite{Plonque2015XIII}.

Another example was the apparent conflict between the primordial
tensor amplitudes inferred from Planck's CMB temperature anisotropies and from the BICEP2
$B$-polarisation measurement.  In this case, an increase to the universe's dark radiation
content~\cite{Archidiacono:2014apa, Giusarma:2014zza} had been proposed as a solution.  However, in the end the
discrepancy was shown to be due to insufficient modelling of the galactic polarised dust emission~\cite{Ade:2015tva}.

While the study of such tensions in {\it actual} data is interesting, for the purpose of this forecast
we prefer not to artificially insert tensions into our mock data;  we shall rather work with unbiased simulated observables, assuming that systematic issues in the
measurements themselves will have been sorted out by the time the {\sc Euclid} mission produces results.

\section{Numerical results}
\label{sec:numericalresults}

\subsection{Unknown neutrino mass}\label{sec:unknown}

\begin{table}
\begin{center}
\resizebox{1\textwidth}{!}{
\begin{tabular}{|l|c|c|c|c|c|}
\hline
\multicolumn{6}{|c|}{\mystrut all} \\
\hline\mystrut
 & $m_{\nu{\rm,fid}}=0.06~{\rm eV}$ & $m_{\nu{\rm,fid}}=0.11~{\rm eV}$ & $m_{\nu{\rm,fid}}=0.3~{\rm eV}$ &
 $m_{\nu{\rm,fid}}=0.7~{\rm eV}$ & $m_{\nu{\rm,fid}}=1~{\rm eV}$ \\
\hline\mystrut
$m_{a{\rm,fid}}=0.01~{\rm eV}$&$0\,-\,0.136$&$0\,-\,0.136$&
$0\,-\,0.144$&$0\,-\,0.137$&$0\,-\,0.144$\\
\hline\mystrut
$m_{a{\rm,fid}}=0.1~{\rm eV}$&$0\,-\,0.144$&$0\,-\,0.144$&
$0\,-\,0.144$&$0\,-\,0.137$&$0\,-\,0.144$\\
\hline\mystrut
$m_{a{\rm,fid}}=0.15~{\rm eV}$&$0.145\,-\,0.256$&$0.145\,-\,0.281$&
$0.145\,-\,0.286$&$0.145\,-\,0.303$&$0.145\,-\,0.317$\\
\hline\mystrut
$m_{a{\rm,fid}}=0.2~{\rm eV}$&$0.155\,-\,0.302$&$0.155\,-\,0.318$&
$0.158\,-\,0.352$&$0.154\,-\,0.331$&$0.155\,-\,0.359$\\
\hline\mystrut
$m_{a{\rm,fid}}=0.5~{\rm eV}$&$0.415\,-\,0.570$&$0.400\,-\,0.567$&
$0.383\,-\,0.566$&$0.392\,-\,0.569$&$0.341\,-\,0.581$\\
\hline
\multicolumn{6}{|c|}{\mystrut ccl}\\
\hline\mystrut
& $m_{\nu{\rm,fid}}=0.06~{\rm eV}$ & $m_{\nu{\rm,fid}}=0.11~{\rm eV}$ & $m_{\nu{\rm,fid}}=0.3~{\rm eV}$ &
$m_{\nu{\rm,fid}}=0.7~{\rm eV}$ & $m_{\nu{\rm,fid}}=1~{\rm eV}$ \\
\hline\mystrut
$m_{a{\rm,fid}}=0.01~{\rm eV}$&$0\,-\,0.144$&$0\,-\,0.144$&
$0\,-\,0.144$&$0\,-\,0.135$&$0\,-\,0.144$\\
\hline\mystrut
$m_{a{\rm,fid}}=0.1~{\rm eV}$&$0\,-\,0.136$&$0\,-\,0.138$&
$0\,-\,0.137$&$0\,-\,0.137$&$0\,-\,0.137$\\
\hline\mystrut
$m_{a{\rm,fid}}=0.15~{\rm eV}$&$0.145\,-\,0.392$&$0.145\,-\,0.507$&
$0.145\,-\,0.749$&$0.145\,-\,0.706$&$0.145\,-\,0.893$\\
\hline\mystrut
$m_{a{\rm,fid}}=0.2~{\rm eV}$&$0.157\,-\,0.407$&$0.161\,-\,0.531$&
$0.168\,-\,0.825$&$0.169\,-\,0.869$&$0.173\,-\,0.902$\\
\hline\mystrut
$m_{a{\rm,fid}}=0.5~{\rm eV}$&$0.233\,-\,0.642$&$0.204\,-\,0.765$&
$0.266\,-\,0.933$&$0.226\,-\,1.033$&$0.253\,-\,1.107$\\
\hline
\multicolumn{6}{|c|}{\mystrut csgx} \\
\hline\mystrut
& $m_{\nu{\rm,fid}}=0.06~{\rm eV}$ & $m_{\nu{\rm,fid}}=0.11~{\rm eV}$ & $m_{\nu{\rm,fid}}=0.3~{\rm eV}$ &
$m_{\nu{\rm,fid}}=0.7~{\rm eV}$ & $m_{\nu{\rm,fid}}=1~{\rm eV}$ \\
\hline\mystrut
$m_{a{\rm,fid}}=0.01~{\rm eV}$&$0\,-\,0.137$&$0\,-\,0.137$&
$0\,-\,0.144$&$0\,-\,0.136$&$0\,-\,0.137$\\
\hline\mystrut
$m_{a{\rm,fid}}=0.1~{\rm eV}$&$0\,-\,0.144$&$0\,-\,0.136$&
$0\,-\,0.144$&$0\,-\,0.137$&$0\,-\,0.137$\\
\hline\mystrut
$m_{a{\rm,fid}}=0.15~{\rm eV}$&$0.145\,-\,0.309$&$0.145\,-\,0.309$&
$0.145\,-\,0.355$&$0.145\,-\,0.331$&$0.145\,-\,0.354$\\
\hline\mystrut
$m_{a{\rm,fid}}=0.2~{\rm eV}$&$0.151\,-\,0.311$&$0.155\,-\,0.330$&
$0.153\,-\,0.387$&$0.154\,-\,0.352036$&$0.157\,-\,0.381$\\
\hline\mystrut
$m_{a{\rm,fid}}=0.5~{\rm eV}$&$0.412\,-\,0.570$&$0.396\,-\,0.567$&
$0.361\,-\,0.566$&$0.375\,-\,0.563$&$0.278\,-\,0.584$\\
\hline
\multicolumn{6}{|c|}{\mystrut cs} \\
\hline\mystrut
& $m_{\nu{\rm,fid}}=0.06~{\rm eV}$ & $m_{\nu{\rm,fid}}=0.11~{\rm eV}$ & $m_{\nu{\rm,fid}}=0.3~{\rm eV}$ &
$m_{\nu{\rm,fid}}=0.7~{\rm eV}$ & $m_{\nu{\rm,fid}}=1~{\rm eV}$ \\
\hline\mystrut
$m_{a{\rm,fid}}=0.01~{\rm eV}$&$0\,-\,0.137$&$0\,-\,0.137$&
$0\,-\,0.144$&$0\,-\,0.144$&$0\,-\,0.144$\\
\hline\mystrut
$m_{a{\rm,fid}}=0.1~{\rm eV}$&$0\,-\,0.144$&$0\,-\,0.137$&
$0\,-\,0.144$&$0\,-\,0.137$&$0\,-\,0.144$\\
\hline\mystrut
$m_{a{\rm,fid}}=0.15~{\rm eV}$&$0.145\,-\,0.321$&$0.145\,-\,0.350$&
$0.145\,-\,0.513$&$0.145\,-\,0.704$&$0.145\,-\,0.824$\\
\hline\mystrut
$m_{a{\rm,fid}}=0.2~{\rm eV}$&$0.152\,-\,0.349$&$0.157\,-\,0.374$&
$0.161\,-\,0.528$&$0.164\,-\,0.738$&$0.166\,-\,0.838$\\
\hline\mystrut
$m_{a{\rm,fid}}=0.5~{\rm eV}$&$0.200\,-\,0.585$&$0.186\,-\,0.604$&
$0.208\,-\,0.711$&$0.187\,-\,0.865$&$0.218\,-\,0.941$\\
\hline
\end{tabular}
}
\caption{1D marginal 95\% credible intervals for the axion mass inferred from 
four different data combinations (indicated above the panels; see text for an explanation of the abbreviations),  assuming different fiducial $m_a$ and $m_\nu$ values.}
\label{tab:allranges}
\end{center}
\end{table}

Table~\ref{tab:allranges} shows the allowed ranges of axion masses for our 25 different combinations of fiducial axion and neutrino
mass values, inferred from four combinations of data set:
\begin{itemize}
\item {\bf ccl}: CMB + clusters,
\item {\bf cs}:  CMB + shear auto-spectrum,
\item {\bf csgx}: CMB + shear auto-correlation + galaxy auto-correlation + shear-galaxy
    cross-correlation, and
\item {\bf all}: CMB + shear auto-correlation + galaxy auto-correlation + shear-galaxy cross correlation
    + clusters.
\end{itemize}

The ``jump effect'' discussed in section~\ref{sec:dectemp} is immediately apparent: for
fiducial axion masses below $m_{\rm QCD}=0.145$~eV, it is not possible to diagnose a nonzero axion mass
even with a future mission such as {\sc Euclid}.  This is in sharp contrast to the case of neutrino masses, where it has been shown that 
even the minimum mass sum of $m_\nu = 0.06$~eV can be detected by {\sc Euclid} with high statistical significance~\cite{Hamann:2012fe}.
As explained earlier, this is because axions with masses smaller than $m_{\rm QCD}$ decouple at temperatures above 
$T_{\rm QCD}=170$~MeV, so that their corresponding present-day number densities suffer from entropy dilution acquired during the QCD phase transition, 
as already illustrated in figure~\ref{fig:na}; in contrast, the present-day neutrino number density is always $n_\nu = 112~{\rm cm}^{-3}$ per flavour, independently of the neutrino mass.
As soon as the fiducial axion mass exceeds $m_{\rm QCD}$, however, a positive detection becomes easily possible.

\begin{figure}[t]
\centering
\includegraphics[scale=0.36]{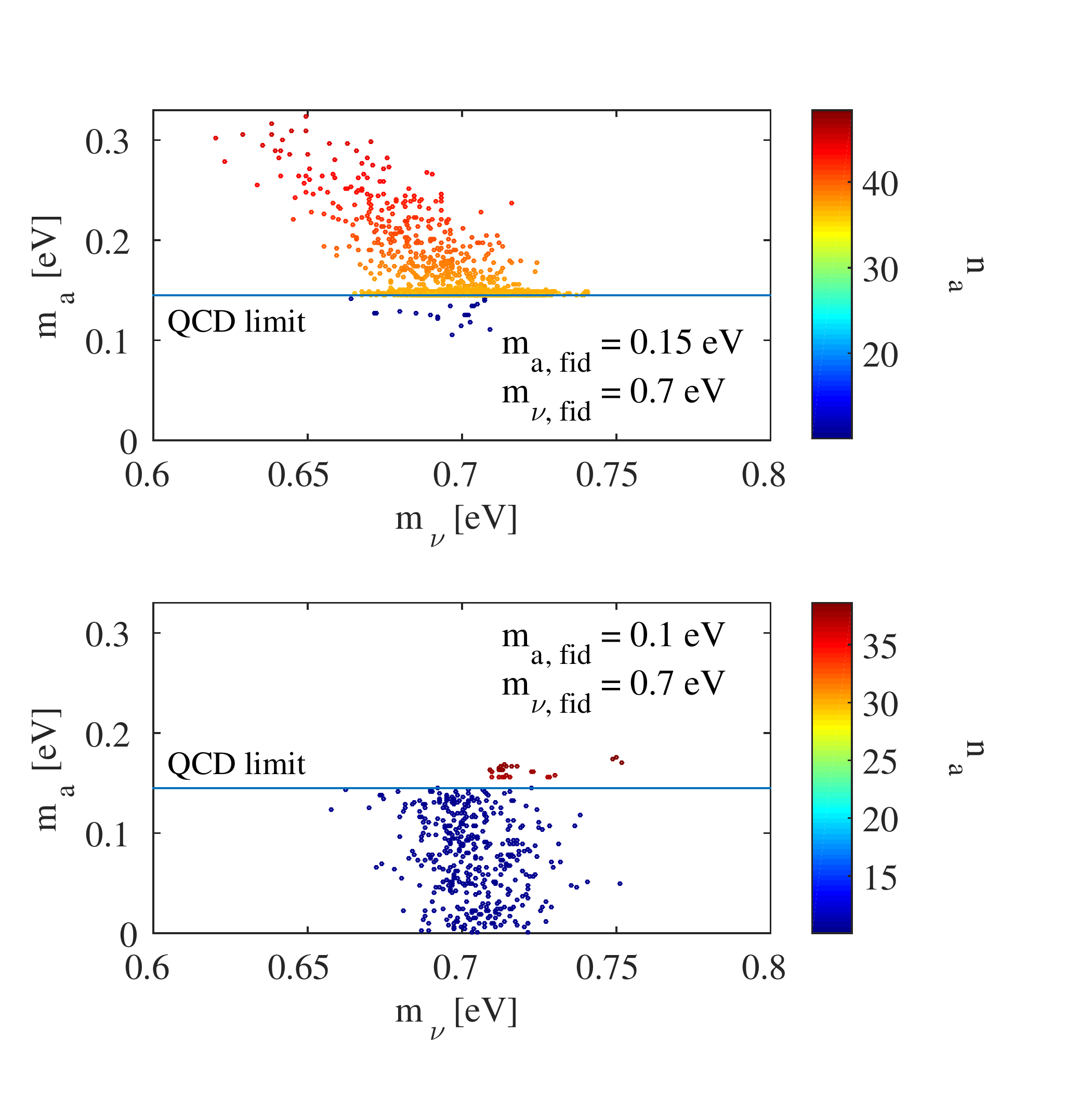}
\includegraphics[scale=0.35]{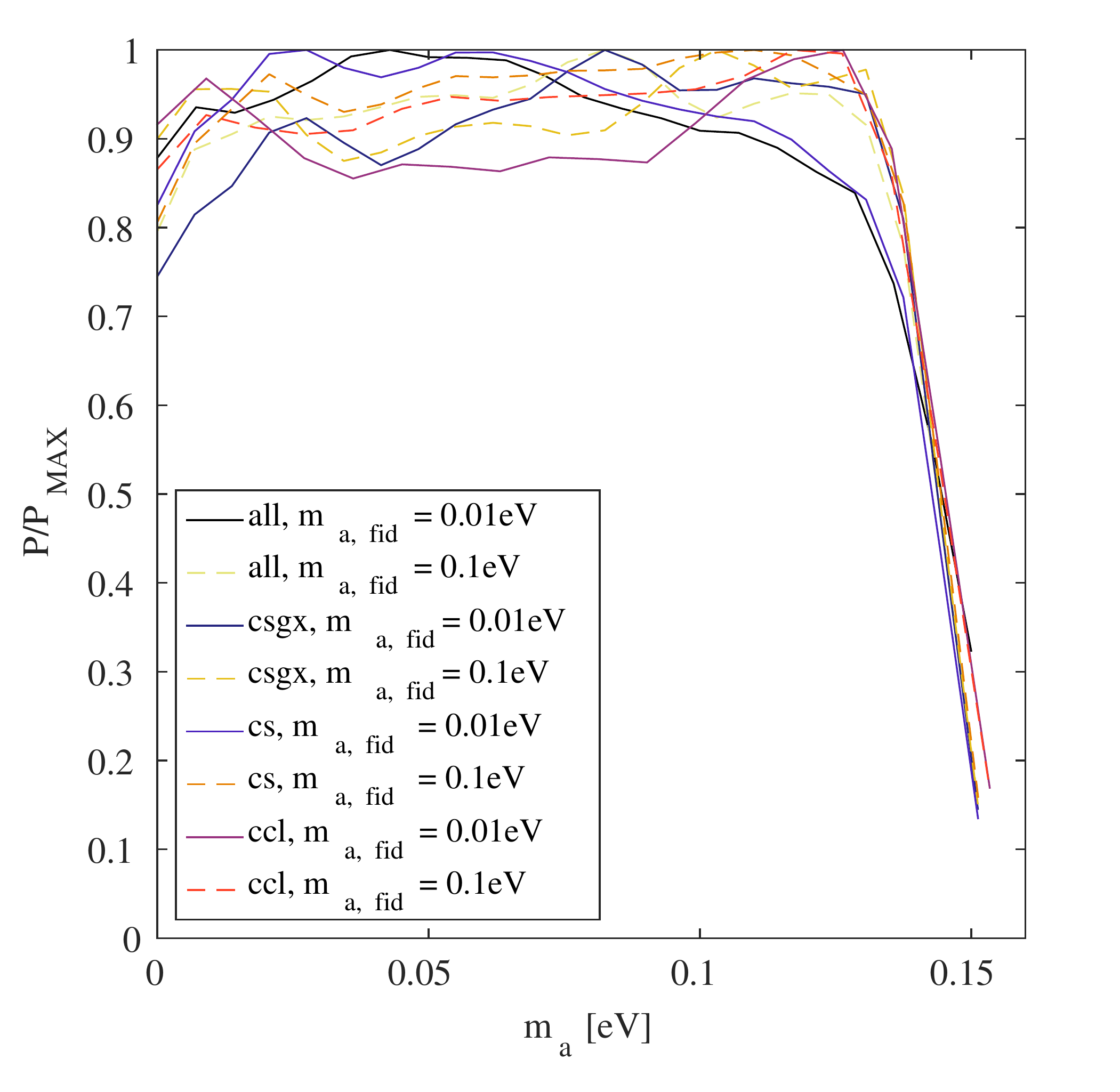}
\caption{{\it Left}: Scatter plot of inferred mass values in the $m_a$-$m_\nu$ plane, assuming a common $m_{\nu{\rm,fid}}=0.7$~eV, 
and  $m_{a{\rm,fid}}=0.15\ {\rm eV} > m_{\rm QCD}$ (top) and $m_{a{\rm,fid}}=0.10\ {\rm eV}<m_{\rm QCD}$ (bottom). The combination of all data sets is considered in both cases.
The colours of the points indicate the corresponding axion number density according to the colour bar. {\it Right}:
1D marginal posterior probability density distribution in  $m_a$ inferred from various combinations of data sets, 
assuming $m_{\nu{\rm,fid}}=0.06\,{\rm eV}$ and two choices of fiducial axion masses below the QCD mass limit.
}
\label{fig:na3D}
\end{figure}

We further illustrate this jump effect in the left panel of figure~\ref{fig:na3D}, which shows two scatter plots of the $m_a$ and $m_\nu$ values from the likelihood analysis of the ``all'' data combination, assuming a common $m_{\nu{\rm,fid}}=0.7$~eV but two different~$m_{a, {\rm fid}}$ values: one just above $m_{\rm QCD}$ in the top panel ($m_{a, {\rm fid}} = 0.15$~eV), and one just below in the bottom ($m_{a,{\rm fid}} = 0.1$~eV).  The axion number density corresponding to each scatter point is denoted by its colour coding.
Clearly, in the top panel, the likelihood analysis returns a high concentration of points at the fiducial axion mass value.  In the usual fashion the rest of the points are scattered around $m_{a, {\rm fid}}$, but with the twist that the density of points is sharply reduced for axion masses below~$m_{\rm QCD}$,  the QCD phase transition mass limit.  

A similar cut-off can also be seen in the bottom panel, except that in this case, because $m_{a, {\rm fid}}< m_{\rm QCD}$, it is those axion masses  exceeding $m_{\rm QCD}$ that are devoid of scatter points, all of which are now spread instead between $0$~eV and $m_{\rm QCD}$.  Indeed, the probability distribution between $m_a = 0$~eV and $m_{\rm QCD}$ is essentially flat in the $m_a$-direction no matter the exact choice of~$m_{a, {\rm fid}}$, as demonstrated in the right panel of figure~\ref{fig:na3D} by the 1D marginal posteriors  for a selection of  $m_{a,{\rm fid}}$ values.  This flatness signifies that the $m_{a, {\rm fid}}< m_{\rm QCD}$ cases are for all purposes indistinguishable from one another.

\begin{figure}[t]
\centering
\includegraphics[width=0.9\textwidth]{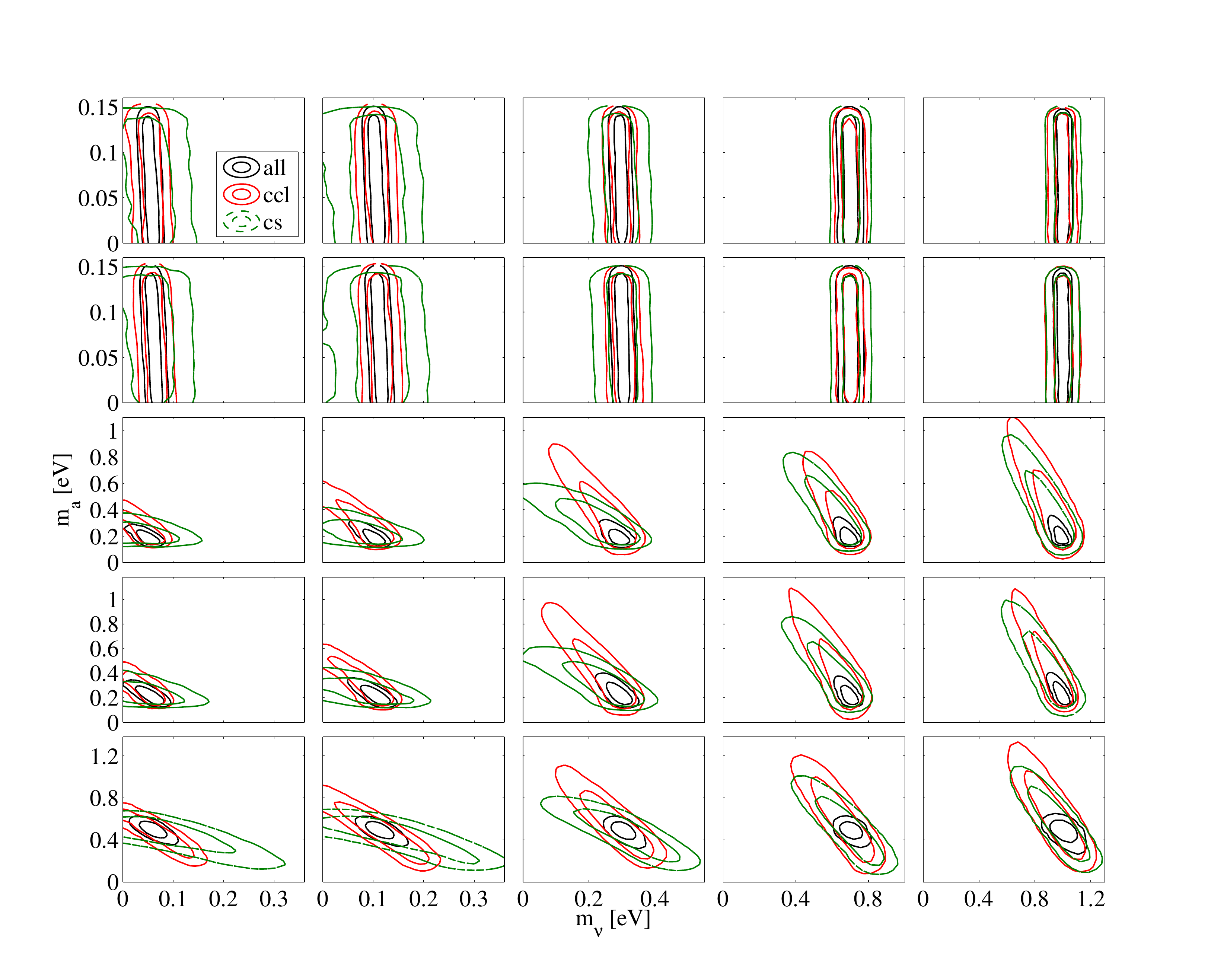}
\caption{2D marginal 68\% and 95\% credible contours for $m_a$ and $m_\nu$ derived from the ``ccl'' (red), ``cs'' (green), and ``all'' (black) 
data combinations. The matrix of fiducial mass combinations is the same as in table~\ref{tab:allranges}.}
\label{fig:2D}
\end{figure}

\begin{figure}[t]
\centering
\begin{tabular}{cc}
\includegraphics[scale=0.37]{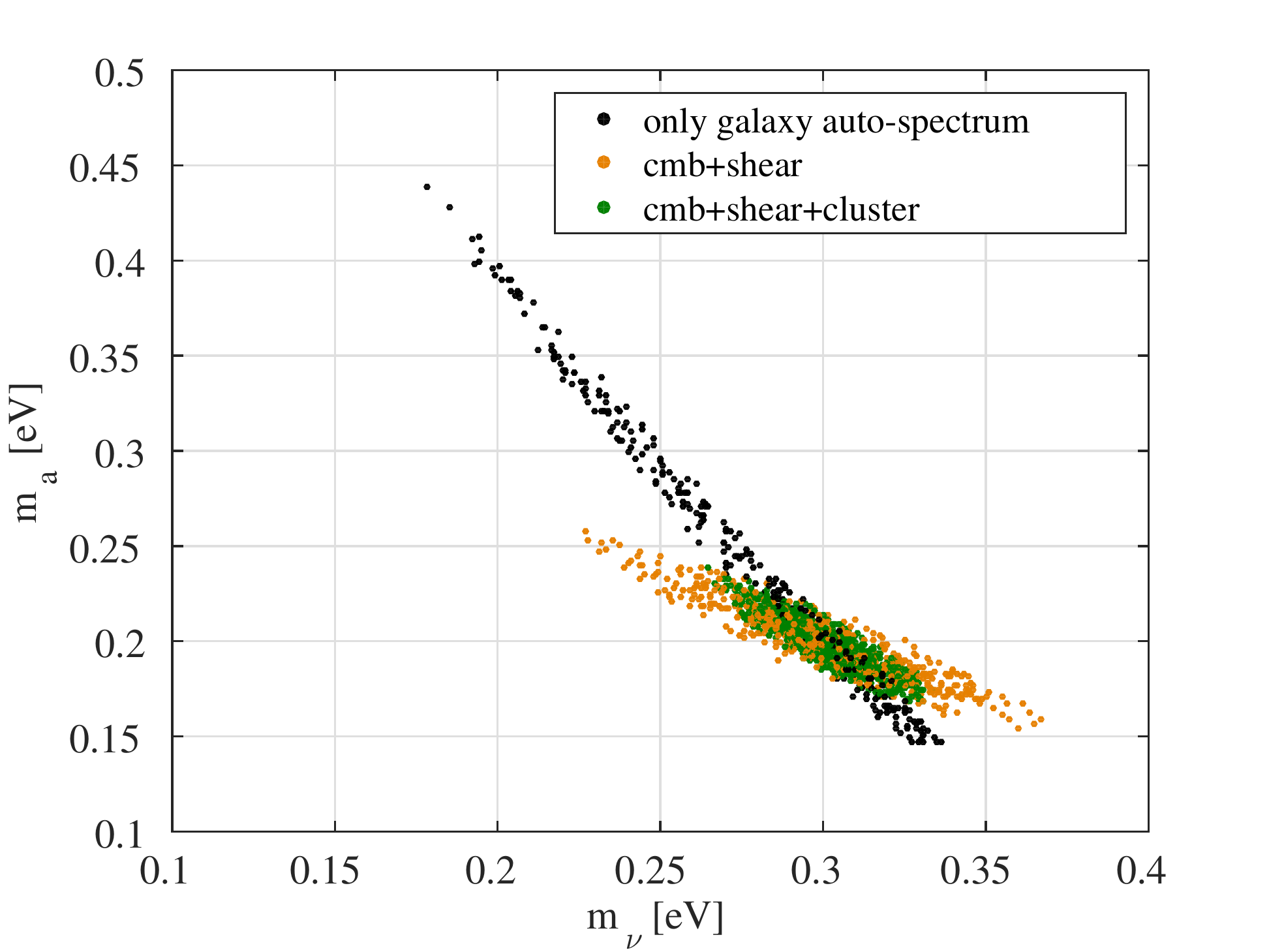}&
\includegraphics[scale=0.37]{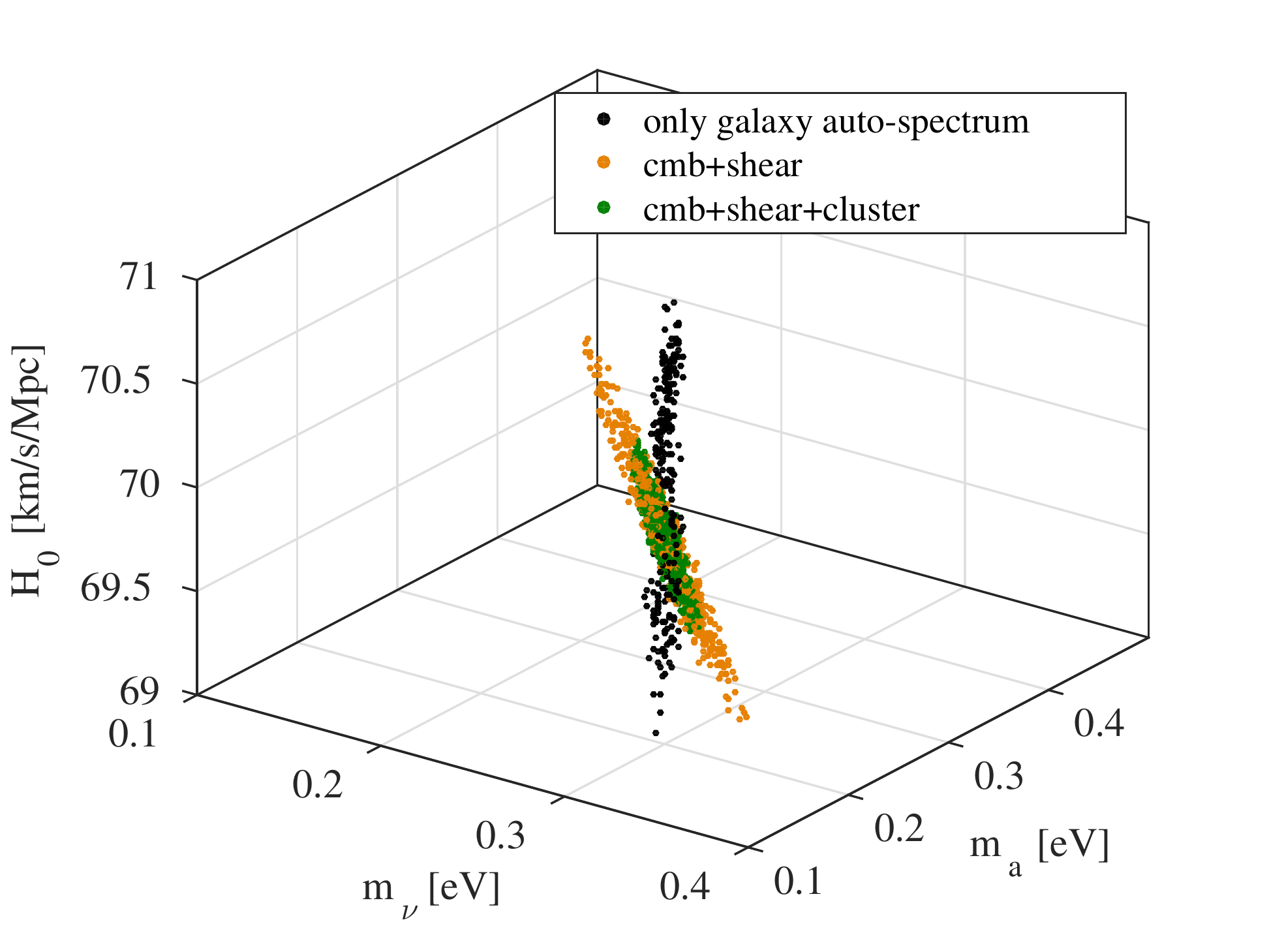}
\end{tabular}
\caption{Scatter plot in the 3D $(H_0,m_a,m_\nu)$-parameter space for the fiducial mass combination $m_{a,{\rm fid}} = 0.2$~eV and $m_{\nu,{\rm fid}}= 0.3$~eV.
 The orange points denote the ``cs'' data combination, the green points  ``cs+clusters'', and the black points the galaxy auto-spectrum. The MCMC fitting has been performed only in this 3D subspace, while all other parameters have been held fixed at their fiducial values. For improved readability we plot only  those points satisfying $\Delta \chi^2 \leq 2$.}
\label{fig:h0}
\end{figure}

Lastly, figure~\ref{fig:2D} shows the 68\% and 95\% contours  in the 2D $(m_a, m_\nu)$-parameter space
derived from various data combinations and for the full matrix of 
 fiducial mass values shown in table~\ref{tab:allranges}.   
 For most fiducial mass combinations there is a strong degeneracy between $m_\nu$ and $m_a$ when the ``ccl'' and
``cs'' data combinations are considered separately.  
This degeneracy is however almost entirely broken for the ``all'' data combination, where the galaxy auto-spectrum is also fitted along with shear and cluster data.
 To understand why this happens, consider the scatter plot in the 3D $(H_0,m_a,m_\nu)$-parameter space displayed in  figure~\ref{fig:h0}.
  Evidently,  the degeneracy directions resulting from the 	``cs''  and ``cs+clusters'' data combinations are close to orthogonal to that from the galaxy auto-spectrum alone.
 Thus, adding the galaxy auto-spectrum not only breaks the degeneracy between $m_a$ and
$m_\nu$, but also the two parameters' respective degeneracies with $H_0$.

\subsection{Fixed neutrino mass}\label{sec:fixed}

Cosmological data of the accuracy studied here will not be available until about a decade from now.  It is
therefore conceivable that the neutrino mass will have been pinned down before then, via, e.g., 
measurements of tritium decay by the KATRIN experiment or by a post-KATRIN generation
experiment~\cite{Formaggio:2014ppa}. We therefore study also the extreme case in which the neutrino mass is
assumed to be exactly known, in order to assess the maximum possible effect our prior knowledge of neutrino masses has
on the cosmological sensitivity to axion masses.
We consider the same 25 combinations  of fiducial axion and neutrino masses, but vary now only the axion mass as a fit
parameter. The likelihood analysis is restricted to the ``all'' combination of data sets.

Table~\ref{tab:mnufixedranges} shows the allowed $m_a$ ranges for all 25 combinations of $m_{a,{\rm fid}}$ and $m_{\nu, {\rm fid}}$.  Comparing with their counterparts in table~\ref{tab:allranges}, 
we can immediately  make the following observations:  On the one hand, for fiducial axion masses below the QCD mass limit, what happens to the neutrino mass continues to have virtually no impact on the sensitivity to $m_a$.    On the other hand, in those cases where $m_{a, {\rm fid}}$  exceeds  $m_{\rm QCD}$, the corresponding error bands  shrink by $\sim 30\%$ to $\sim 60\%$ upon fixing the neutrino mass.

\begin{table}[t]
\begin{center}
\resizebox{1\textwidth}{!}{
\begin{tabular}{|l|c|c|c|c|c|}
\hline
\multicolumn{6}{|c|}{\mystrut all, fixed neutrino masses} \\
\hline\mystrut
 & $m_{\nu{\rm,fid}}=0.06~{\rm eV}$ & $m_{\nu{\rm,fid}}=0.11~{\rm eV}$ & $m_{\nu{\rm,fid}}=0.3~{\rm eV}$ &
 $m_{\nu{\rm,fid}}=0.7~{\rm eV}$ & $m_{\nu{\rm,fid}}=1~{\rm eV}$ \\[0.1cm]
\hline\mystrut
$m_{a{\rm,fid}}=0.01~{\rm eV}$&$0\,-\,0.135$&$0\,-\,0.134$&
$0\,-\,0.136$&$0\,-\,0.137$&$0\,-\,0.137$\\
\hline\mystrut
$m_{a{\rm,fid}}=0.1~{\rm eV}$&$0\,-\,0.138$&$0\,-\,0.137$&
$0\,-\,0.137$&$0\,-\,0.144$&$0\,-\,0.137$\\
\hline\mystrut
$m_{a{\rm,fid}}=0.15~{\rm eV}$&$0.140\,-\,0.211$&$0.140\,-\,0.213$&
$0.140\,-\,0.212$&$0.140\,-\,0.215$&$0.140\,-\,0.223$\\
\hline\mystrut
$m_{a{\rm,fid}}=0.2~{\rm eV}$&$0.159\,-\,0.250$&$0.158\,-\,0.253$&
$0.157\,-\,0.253$&$0.157\,-\,0.256$&$0.154\,-\,0.256$\\
\hline\mystrut
$m_{a{\rm,fid}}=0.5~{\rm eV}$&$0.454\,-\,0.544$&$0.453\,-\,0.545$&
$0.454\,-\,0.544$&$0.455\,-\,0.543$&$0.454\,-\,0.543$\\
\hline
\end{tabular}
}
\caption{1D marginal 95\% credible intervals for the axion mass inferred from the ``all'' data combination, assuming different fiducial $m_a$ and $m_\nu$ values.
In each case the neutrino mass has been fixed at the fiducial value in the analysis.}
\label{tab:mnufixedranges}
\end{center}
\end{table}

These results can be understood from an inspection of the 2D marginal credible contours in figure~\ref{fig:2D}.  When the neutrino mass is left as a free parameter, construction of the 1D marginal posterior for the axion mass $m_a$ consists essentially of integrating the 2D ellipse along the $m_\nu$-direction.   With the neutrino mass fixed, however, the 1D marginal posterior for $m_a$ is simply the cross-section of the ellipse at $m_\nu=m_{\nu, {\rm fid}}$.   Thus, because the ellipses in the $m_{a, {\rm fid}}< m_{\rm QCD}$ cases are almost exactly aligned with the $m_a$-axis, the two procedures yield very similar results.  In contrast, the ellipses in the $m_{a, {\rm fid}}> m_{\rm QCD}$ scenarios are all inclined at an angle;  assuming that the ellipses represent perfect 2D Gaussian distributions, integration along the $m_\nu$-direction will always yield a 1D posterior for $m_a$ that is broader than the cross-section at $m_\nu = m_{\nu, {\rm fid}}$, with the greatest disparity occurring when the inclination approaches $45^\circ$.

\subsection{When can a detection be claimed?}

Having calculated the formal sensitivity to $m_a$ for a variety of different fiducial models we might now also
ask the question: How strongly would a model with $m_a=0$ be disfavoured depending on the assumed fiducial
model?

Table~\ref{tab:chi2} shows the $\Delta \chi^2$ values for the best-fit model assuming $m_a=0$,
relative to the global best fit (i.e., the fiducial model), derived from the ``all'' data set  for the 25 fiducial mass combinations.  (Note that we have also varied the neutrino mass as a fit parameter.)
Looking at the first two rows,  it is immediately clear 
that fiducial axion masses below 0.15~eV, for which $\Delta \chi^2 \simeq 0$, 
will not in practice be distinguishable from zero given the accuracy of the data used here.
This conclusion is consistent with our understanding of the Bayesian results presented in section~\ref{sec:unknown} and~\ref{sec:fixed}.

As soon as the fiducial axion mass surpasses 0.15~eV, the $\Delta \chi^2$ value jumps to the range 13--37, depending on the 
fiducial neutrino mass; the larger the $m_{\nu,{\rm fid}}$ value, the lower the~$\Delta \chi$.  Physically, this trend can be traced to a very small residual degeneracy between $m_a$ and $m_\nu$,
so that a  1~eV neutrino could partially fill in the role of a 0.15~eV axion while a 0.06~eV neutrino could not.  But this is ultimately only an academic point of interest, because
$\Delta \chi^2$ values as large as  13--37 at the cost of just one additional fit parameter would in any case constitute ``strong'' evidence in favour of a nonzero axion mass, no matter whether we apply the Bayes information criterion
(BIC), the Akaike information criterion (AIC),  or other constructs.

 We therefore conclude that for any axion mass above the mass threshold $0.15$~eV required for the axion to decouple after the
QCD epoch, Planck data in combination with a {\sc Euclid}-like photometric
survey will have sufficient accuracy  to reliably measure the axion mass.   If on the other hand the real axion mass falls below the QCD mass threshold, 
then the corresponding axion number density is so low that a future survey with a much better accuracy would be needed to further tighten the axion mass bound from above.

\begin{table}[t]
\begin{center}
\resizebox{1\textwidth}{!}{
\begin{tabular}{|l|c|c|c|c|c|}
\hline\mystrut
 & $m_{\nu{\rm,fid}}=0.06~{\rm eV}$ & $m_{\nu{\rm,fid}}=0.11~{\rm eV}$ & $m_{\nu{\rm,fid}}=0.3~{\rm eV}$ &
 $m_{\nu{\rm,fid}}=0.7~{\rm eV}$ & $m_{\nu{\rm,fid}}=1~{\rm eV}$ \\
\hline\mystrut
$m_{a{\rm,fid}}=0.01~{\rm eV}$&$0$&$0$& 
$0$& $0$ & $0$ \\
\hline\mystrut
$m_{a{\rm,fid}}=0.1~{\rm eV}$&$0$&$0$&
$0$&$0$&$0$\\
\hline\mystrut
$m_{a{\rm,fid}}=0.15~{\rm eV}$&$37$&$36$&
$31$&$18$&$13$\\
\hline\mystrut
$m_{a{\rm,fid}}=0.2~{\rm eV}$&$47$&$46$&
$40$&$25$&$18$\\
\hline\mystrut
$m_{a{\rm,fid}}=0.5~{\rm eV}$&$75$&$71$&
$64$&$55$&$38$\\
\hline
\end{tabular}
}
\caption{$\Delta \chi^2$ values (rounded to the nearest whole number) for the best-fit model assuming $m_a$ fixed at zero, relative to the global best-fit, derived from the ``all'' data set for the 25 fiducial mass combinations.}
\label{tab:chi2}
\end{center}
\end{table}

%
%\begin{table}[t]
%\begin{center}
%\resizebox{1\textwidth}{!}{
%\begin{tabular}{|l|c|c|c|c|c|}
%\hline\mystrut
% & $m_{\nu{\rm,fid}}=0.06~{\rm eV}$ & $m_{\nu{\rm,fid}}=0.11~{\rm eV}$ & $m_{\nu{\rm,fid}}=0.3~{\rm eV}$ &
% $m_{\nu{\rm,fid}}=0.7~{\rm eV}$ & $m_{\nu{\rm,fid}}=1~{\rm eV}$ \\
%\hline\mystrut
%$m_{a{\rm,fid}}=0.01~{\rm eV}$&$0.0657$&$0.0118$& 
%$0.0823$&$-0.128$&$0.123$\\
%\hline\mystrut
%$m_{a{\rm,fid}}=0.1~{\rm eV}$&$-0.0868$&$0.200$&
%$0.311$&$0.217$&$0.339$\\
%\hline\mystrut
%$m_{a{\rm,fid}}=0.15~{\rm eV}$&$36.650$&$36.281$&
%$30.866$&$18.120$&$13.121$\\
%\hline\mystrut
%$m_{a{\rm,fid}}=0.2~{\rm eV}$&$47.046$&$45.728$&
%$39.585$&$25.032$&$17.680$\\
%\hline\mystrut
%$m_{a{\rm,fid}}=0.5~{\rm eV}$&$75.133$&$70.587$&
%$63.859$&$54.976$&$37.553$\\
%\hline
%\end{tabular}
%}
%\caption{$\Delta \chi^2$ values for the best-fit model assuming $m_a$ fixed at zero, relative to the global best-fit, derived from the ``all'' data set for the 25 fiducial mass combinations.}
%\label{tab:chi2}
%\end{center}
%\end{table}

\section{Conclusions}
\label{sec:discussion}

The next generation of large-volume photometric surveys, notably {\sc Euclid}, together with CMB data from the Planck mission are almost guaranteed to reach a sensitivity to the cosmic hot dark matter content good enough to measure the sum of neutrino masses $\sum m_\nu$  with high significance.  This is so even if $\sum m_\nu$ should  take on its 
most pessimistic value, 60~meV, the minimum  established by  flavour oscillation experiments. 
Complementarily, laboratory experiments will provide a direct measurement of the effective electron neutrino mass,  albeit likely on a longer timescale.
Depending on what is actually found at these experiments, there will be a new frontier of either identifying an additional hot dark matter component, or placing very precise constraints on non-neutrino contributions, e.g.,  in the form of a hot axion population.

Motivated by the IAXO proposal for a next-generation solar axion search, we have explored the range of axion masses that is accessible to cosmological hot dark matter searches. 
For large enough axion masses, thermal axions are produced after the QCD epoch by very efficient axion-pion interactions, so that the resulting axion population is almost comparable to  that of one species of neutrinos.
For small masses, however, axion freeze-out occurs before  the QCD epoch, where axion production from interactions with free quarks and gluons is much less efficient, and 
the final axion population is invariably diluted by the large number of colour degrees of freedom that disappear when the temperature of the universe  drops below~$T_{\rm QCD}$.

Based on this understanding, and assuming that $T_{\rm QCD}  = 170$~MeV, we find that axion masses larger than $\sim 0.15$~eV can be easily pinned down by cosmology.  If on the other hand the axion mass is lower than this threshold, then cosmological observations are essentially blind to the hot  axion population for the foreseeable future.

%When optimising the IAXO apparatus and search strategies, we conclude that the mass range above 0.15~eV would seem to have lesser priority. Accessing this mass range with IAXO requires helium filling of the conversion region to achieve a match between the axion search mass and the refractive photon mass, a method that has been used at CAST and requires sequential pressure settings to cover a broad range of axion search masses. Therefore, cosmology can help to focus the IAXO search on the most interesting parameter range.  {\color {red} (YW: I don't quite get the sentiment of this sentence. Georg?)}

When optimising the IAXO apparatus and search strategies, we
conclude that the mass range $m_a\lesssim0.15~{\rm eV}$ should have the
highest priority because larger axion masses can be constrained by
cosmology. To study this higher mass range with IAXO requires
helium filling of the conversion region to achieve a match between
the axion search mass and the refractive photon mass, a method that
has been used at CAST and requires sequential pressure settings to
cover a broad range of axion search masses. Therefore, cosmology
can help to simplify the IAXO search strategy in that the
technically most difficult mass range may well have been excluded
by cosmology before IAXO becomes operational.

\section*{Acknowledgments}

We acknowledge use of computing resources from the Danish Center for Scientific Computing
(DCSC), and partial support by the Deutsche Forschungsgemeinschaft through grant
No.\ EXC 153 (Excellence Cluster ``Universe'') and by the European Union through the Initial Training Network
``Invisibles,'' grant No.\ PITN-GA-2011-289442.

\bibliographystyle{utcaps}
\providecommand{\href}[2]{#2}\begingroup\raggedright
\endgroup

\end{document}